\documentclass[sigconf]{acmart}
\acmBooktitle{ }
\usepackage{soul}
\usepackage{braket}
\usepackage{url}
\usepackage{caption}
\usepackage{subcaption}
\usepackage{amsmath}
\usepackage{amsthm}
\usepackage{booktabs}
\usepackage{multirow}
\usepackage{multicol}
\usepackage{makecell}
\usepackage{pifont}
\usepackage[linesnumbered,ruled,vlined]{algorithm2e}
\usepackage[mathscr]{eucal}
\SetKwRepeat{Do}{do}{while}
\let\oldnl\nl
\newcommand{\nonl}{\renewcommand{\nl}{\let\nl\oldnl}}
\newtheorem{theorem}{Theorem}
\newtheorem{corollary}{Corollary}[theorem]

\usepackage{algpseudocode}%
\usepackage{tikz}
\usepackage{pifont}

\renewcommand\footnotetextcopyrightpermission[1]{}
\newcommand*\circled[1]{\tikz[baseline=(char.base)]{
            \node[shape=circle,draw,inner sep=1pt] (char) {#1};}}
\begin{document}
\title[NAC-QFL: Noise Aware Clustered Quantum Federated Learning]{NAC-QFL: Noise Aware Clustered Quantum Federated Learning \vspace{-0.25cm}}
\author{Himanshu Sahu}
\email{himanshusahu.rs.cse21@iitbhu.ac.in}
\affiliation{%
  \institution{IIT (BHU) Varanasi, India}
  \country{India}
  \vspace{-0.2cm}
}

\author{Hari Prabhat Gupta}
\email{hariprabhat.cse@iitbhu.ac.in}
\affiliation{%
  \institution{IIT (BHU) Varanasi, India}
  \country{India}
}

\renewcommand{\shortauthors}{Himanshu et al.}

\begin{abstract}
Recent advancements in quantum computing, alongside successful deployments of quantum communication, hold promises for revolutionizing mobile networks. While Quantum Machine Learning (QML) presents opportunities, it contends with challenges like noise in quantum devices and scalability. Furthermore, the high cost of quantum communication constrains the practical application of QML in real-world scenarios. This paper introduces a noise-aware clustered quantum federated learning system that addresses noise mitigation, limited quantum device capacity, and high quantum communication costs in distributed QML. It employs noise modelling and clustering to select devices with minimal noise and distribute QML tasks efficiently. Using circuit partitioning to deploy smaller models on low-noise devices and aggregating similar devices, the system enhances distributed QML performance and reduces communication costs. Leveraging circuit cutting, QML techniques are more effective for smaller circuit sizes and fidelity. We conduct experimental evaluations to assess the performance of the proposed system. Additionally, we introduce a noisy dataset for QML to demonstrate the impact of noise on proposed accuracy. 
 \end{abstract}

\vspace{-0.2cm}

\vspace{-0.3cm}

\maketitle
\vspace{-0.3cm}

\section{Introduction}
Quantum computing along with the successful deployment of quantum communication promise to revolutionize mobile networks~\cite{book2,qmlmobile,qleap}. For example, in a smart city context, quantum computing can process vast amounts of real-time data collected from various sources such as traffic sensors, public transportation systems, and weather forecasts. This data can be efficiently analyzed on quantum devices to optimize traffic flow, reduce congestion, and enhance overall urban mobility. Quantum Machine Learning (QML) is one of the key areas in quantum research that is continuously growing with the successful implementation of quantum counterparts~\cite{qml,book2}.

Near-term quantum devices are \textbf{N}oisy (i.e. Instability of quantum states) and \textbf{I}ntermediate \textbf{S}cale (i.e limited number of qubits) \textbf{Q}uantum device termed as \textbf{NISQ}-quantum computers. Quantum noise is inevitable due to system-environment interaction and qubit decoherence~\cite{book}. Small changes in temperature or stray electrical or magnetic fields can disturb the state of qubits. Scalability is the other challenge in quantum computing and increasing the capacity directly without fault tolerance resulting more noise in the circuit execution. 
Due to system capacity and noise, QML model training on real hardware is limited to toy examples. The Quantum Error Correction (QEC) can only be implemented by adding redundancy which requires additional qubit resources. So, QEC is infeasible with the limited capacity of NISQ-era devices. It would be better to adopt an alternate solution, Quantum Error Mitigation (QEM). However, QEM only attempts to produce accurate expectation values for an observable. Other than noise scalability can be addressed using distributed computing that can train a QML model by coherently utilizing multiple quantum devices and quantum communication. Centralized QML training suffers from additional issues of data privacy and communication overhead. Transferring quantum encoded data and performing remote operations incurs huge communication overhead. In contrast, distributed setup has several advances such as it will help in noise mitigation inherently due to circuit partitioning. 

Considering loosely coupled distributed systems and privacy concerns, Quantum Federated Learning (QFL) comes out as a better approach that ensures data privacy and communication efficiency~\cite{qfl}~\cite{qsa}. QFL enables collaborative learning across decentralized devices without sharing raw data in mobile networks and mobile computing. Quantum devices quickly process the data, thereby preserving privacy while still achieving accurate model training. 
It is noteworthy that the cost of per-bit communication is notably high in quantum communication. QFL also aids in reducing data communication costs as it only requires training weights to be transferred. Moreover, QFL is capable of providing better noise mitigation due to the reduced circuit depth~\cite{noisecircuit}. In the QFL scenario, quantum-based authentication protocols can be employed to verify the participant devices. QFL is assumed to be the viable solution that can utilize the near-term NISQ devices for learning deeper QML models to provide faster convergence, communication efficiency,  better accuracy and noise mitigation simultaneously. 

This paper proposes a \textbf{N}oise \textbf{A}ware \textbf{C}lustered \textbf{Q}uantum \textbf{F}ederated \textbf{L}earning system, abbreviated as NAC-QFL system, that considers noise as a key factor during device selection and communication efficient clustering to effectively mitigate the noise and faster convergence in a clustered federated. The NAC-QFL system outlines a procedure for clustering devices for QFL, followed by a noise modelling approach to identify devices with the least noise for participation. Additionally, NAC-QFL partitions the QML model to distribute tasks across clustered devices. This system addresses noise mitigation in quantum devices, limited availability of qubit devices, and high quantum communication costs. By employing noise modelling methods, NAC-QFL reduces noise in QML and improves distributed QML performance by running smaller models on devices with better noise profiles. Leveraging circuit cutting, QML techniques prove more effective for smaller circuit sizes. The clustered architecture facilitates the aggregation of similar devices, lowering the cost of distributed training. Furthermore, clustering reduces communication costs as not all devices need to communicate directly with the aggregation server.

The rest of the paper is organized as follows: Section~\ref {sec:2} discusses the existing work focusing on NISQ-Quantum computers, quantum noise, QML, and QFL. Section~\ref{sec:3} provides the NAC-QFL system design, workflow and methodology. Section~\ref{sec:4} provides the performance evaluation with results and finally, the conclusion is provided in Section~\ref{sec:5}.

\section{Background}
\label{sec:2}
This section introduces existing work focusing on NISQ-Quantum computers, quantum noise, QML, and QFL. It also presents the motivation and major contributions of this work.
\subsection{NISQ-Quantum Computer}
\begin{figure}[t]
    \centering
    \begin{subfigure}[b]{0.4\linewidth}
         \centering
         \includegraphics[scale=.18]{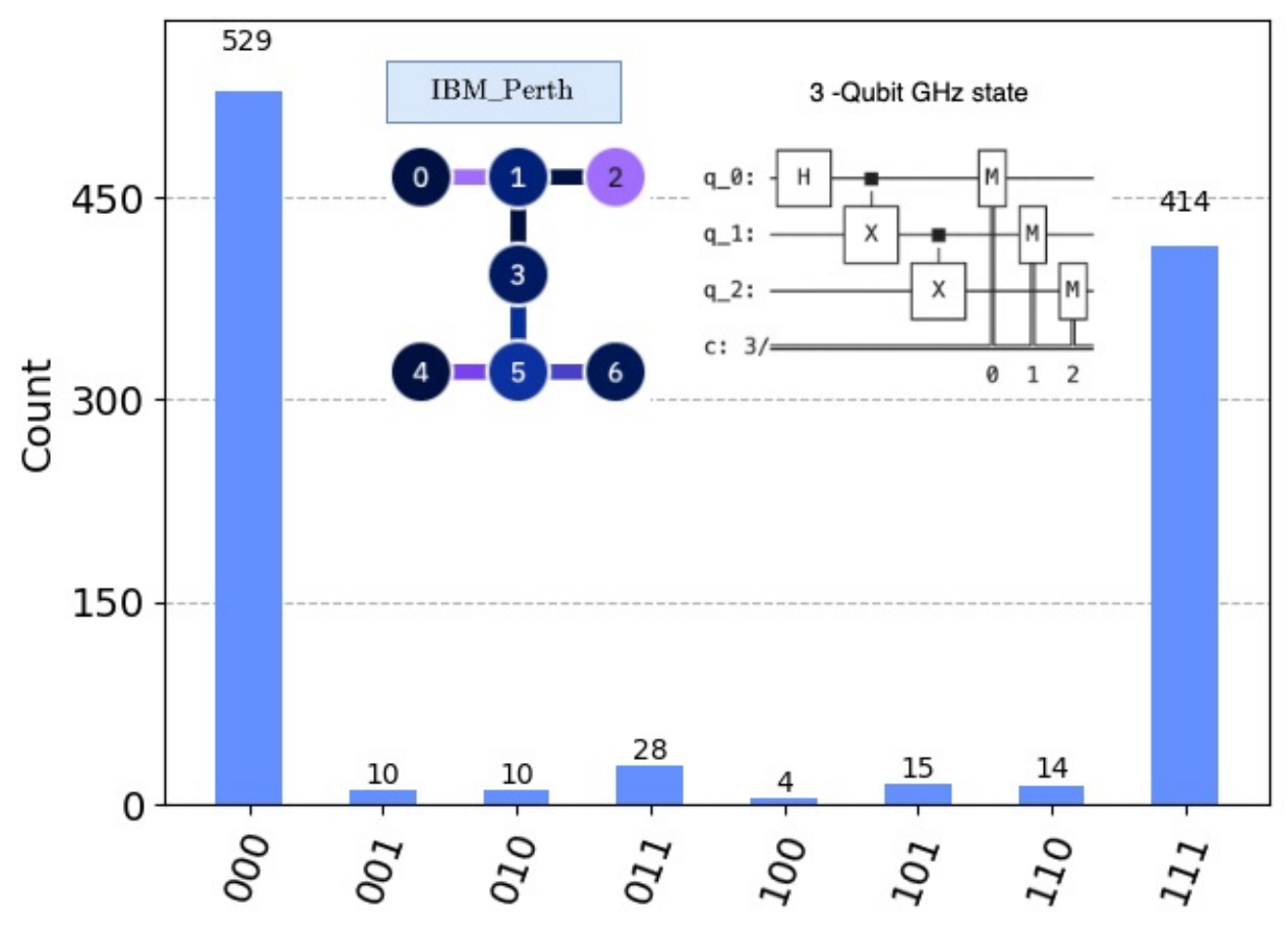}
         \subcaption{\label{fig:fig1a} \footnotesize 3-Qubit GHZ-IBM\_Perth}
     \end{subfigure}
     \hfill
      \begin{subfigure}[b]{0.5\linewidth}
         \includegraphics[scale=.13]{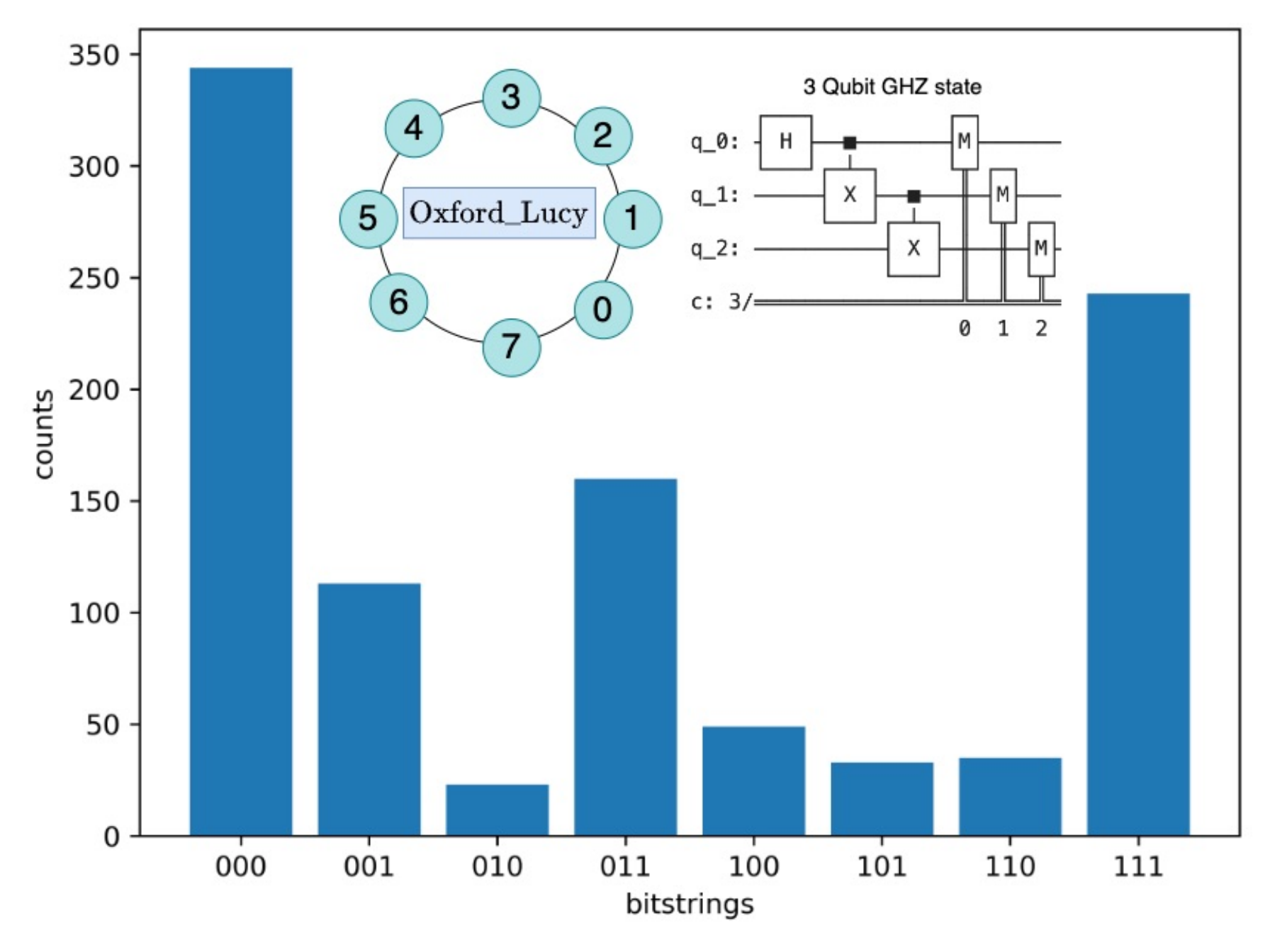}
         \subcaption{\label{fig:fig1b} \footnotesize 3-Qubit GHZ-OQC\_Lucy}
     \end{subfigure}
     \hfill
      \begin{subfigure}[b]{0.45\linewidth}
         \centering
         \includegraphics[scale=.3]{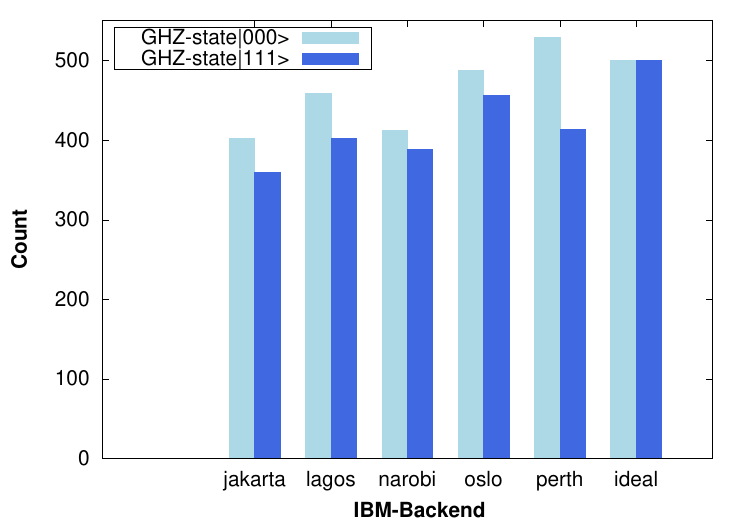}
         \subcaption{ \label{fig:fig1c}\footnotesize 7-Qubit backends}
     \end{subfigure}
     \hfill
      \begin{subfigure}[b]{0.45\linewidth}
         \centering
         \includegraphics[scale=.3]{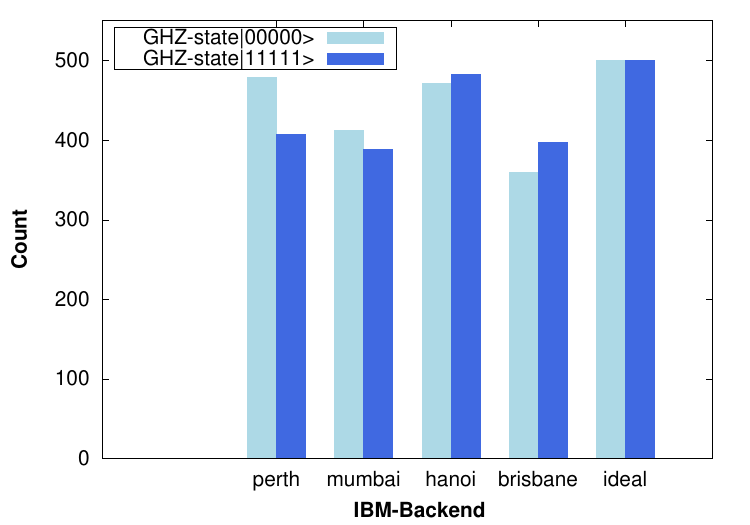}
         \subcaption{ \label{fig:fig1d}\footnotesize 7-127 Qubit backends}
     \end{subfigure}
     \hfill
    \vspace{-0.35cm}
    \caption{\label{fig:fig1} \small Effect of device topology and noise model on the circuit performance. Parts (a)-(b) show the effect of topology on three qubit GHZ circuit. Part (c) shows the effect of a noise model for the same topology seven-qubit IBM device. Part (d) shows the performance variation of five-qubit GHZ performance in different capacity IBM devices.} 
    \vspace{-0.5cm}
\end{figure}

Quantum computers rely on qubit implementation technology. However, such implementations are unstable and remain in a valid quantum state for only a limited time, known as coherence time. Various qubit implementation technologies, including superconducting qubits, neutral atoms, and trapped ions, yield diverse quantum hardware~\cite{book2}. Available hardware varies in architecture, topology, capacity, and logic gates, which are crucial factors in evaluating noise effects during circuit execution. Figure~\ref{fig:fig1} demonstrates the performance variation of due to topology, noise model, and qubit capacity. 
\subsubsection*{Transpilation} A quantum circuit, representing a quantum algorithm, comprises logical qubits and quantum logic gates in an abstract form. A transpiler serves as a tool to translate this abstract circuit into one that is functionally equivalent for a designated quantum computer. It optimizes the circuit to mitigate the impact of noise, decoherence, and errors, while also introducing new gates like swap gates for circuit transformation. However, transpilation can result in increased circuit depth, potentially amplifying noise. Hence, ensuring compatibility of gate sets becomes crucial for optimizing the performance of a quantum algorithm on hardware.
\vspace*{-0.3cm}

\subsection{Quantum Noise}
 Quantum computers are highly susceptible to noise due to the instability of qubits and environment interaction. The quantum noise mainly comes from four major sources: i) Access interface- causes system and environment interaction, ii) Undesired qubit-qubit interaction - causes unwanted mixed states, iii) Imperfect gates - causes deviation from desired evolution, and iv) Leakage - causing decoherence~\cite{qcbench}. 
The density matrix is used for the evolution of the quantum state in the presence of noise \cite{book2}. Let \(\rho\) is the initial qubit state then effect of environment noise \(\rho_{env}\) is described as 
 \begin{equation}
 \label{eq:1}
    \mathcal{N}(\rho) = Tr_{\text{env}}[U(\rho \oplus \rho_{\text{env}}) U^{\dagger}].
 \end{equation}
 The operator sum representation of Eq.~\eqref{eq:1} is provided with the help of $n$-pairs of Kraus operators ($\{A_i,A_i^\dag\}_{i=1}^{n}$), \textit{i.e.,}
\begin{equation}
\label{eq:kraus}
    \mathcal{N}(\rho) = \sum_{i=1}^{n}A_i\rho A_i^\dag,
    \vspace{-.2cm}
\end{equation}
\noindent where, $\sum_{i=1}^{n} A_i A_i^\dag = \mathbb{I}$, the identity matrix\cite{book}. Pauli noise is defined as the probability of happening of one of Pauli gates (X, Y, Z) after a quantum operation is performed on a qubit, \textit{i.e.,}
\begin{equation}
\label{eq:pauli}
    \mathcal{N}_P^{Pauli}(\rho) = \sum_{i=1}^{4}A_i\rho A_i^\dag, 
\end{equation}  
where \(A_0\) = \(\sqrt{(1-(p_x+p_y+p_z)}\mathbb{I}\), \(A_1\) = \(\sqrt{p_x}\sigma_x \), \(A_2\) = \(\sqrt{p_y}\sigma_y \) and \(A_3\) = \(\sqrt{p_z}\sigma_z \). The \(\sigma_x, \sigma_y, \sigma_z\) represent Pauli-X,Y, Z matrix as 
\(\sigma_{x} = \begin{bmatrix}
            0&1\\
            1& 0
        \end{bmatrix}\), \(\sigma_{y} = \begin{bmatrix}
            0& -i\\
            i & 0
        \end{bmatrix}\) and \(\sigma_{z} = \begin{bmatrix}
            1&0\\
            0&-1
        \end{bmatrix}\). 

\subsubsection{Noisy Quantum Channel}
\begin{figure}[htbp]
    \centering
    \includegraphics[scale=0.5]{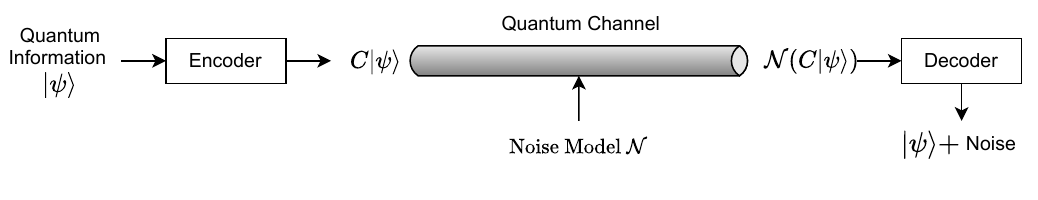}
    \vspace{-0.8cm}
    \caption{\footnotesize Noisy Quantum Channel: Encoding quantum information and transmitting it over a quantum communication channel which adds noise and quantum information is recovered after decoding. }
    \label{fig:fignc}
\end{figure}

Noisy Quantum channels are used to represent noise in quantum communication and quantum computers. The behaviour of the noisy channel is shown in Figure~\ref{fig:fignc}, where a channel coding method encodes/decodes the quantum information.
Different noisy channels are described as follows: \\
$\noindent$ $\bullet$ \textit{Bit Flip Channel} flips the qubit state from \(\ket{0} \mapsto \ket{1}\) (or vice-versa) with \(p\) probability given as 
        \begin{equation}
          \label{eq:bitflip}
            \mathcal{N}_p^{BF}(\rho) = (1-p)\rho + p\sigma_x\rho\sigma_x,
          \end{equation}
        where, \(0 \leq p \leq 1\) and \(\sigma_{x}\) is given in Eq.~\ref{eq:pauli}.\\
       
$\noindent$ $\bullet$ \textit{Phase Flip channel} flips the relative phase of qubit with probability \(p\) is given as
        \begin{equation}
        \label{eq:phaseflip}
            \mathcal{N}_p^{PF}(\rho) = (1-p)\rho + p\sigma_z\rho\sigma_z,
        \end{equation}
        where, \(0 \leq p \leq 1\) and \(\sigma_{z}\) is given in Eq.~\ref{eq:pauli}.\\
$\noindent$ $\bullet$ \textit{Depolarizing Channel} applies all Pauli channels with equal probability \(p\) is given as 
        \begin{equation}
        \label{eq:depo}
            \mathcal{N}_p^{depo}(\rho) = (1-\frac{3}{4}p)\rho + \frac{p}{4}(\sigma_x\rho\sigma_x+\sigma_y\rho\sigma_y+\sigma_z\rho\sigma_z),
        \end{equation}
        where, \(0 \leq p \leq 1\) and \(\sigma_{x},\sigma_{y},\sigma_{z}\) is given in Eq.~\ref{eq:pauli}.\\
 
$\noindent$ $\bullet$ \textit{Amplitude damping channel} causes decoherence of qubit excited state due to continuous emission, \textit{i.e.,}
    \begin{equation}
    \label{eq:ampdamp}
        \mathcal{N}^{damp}_p(\rho) = K_0\rho K_0^\dag + K_1\rho K_1^\dag,
    \end{equation}
    where, $K_0 = \begin{bmatrix}
            1&0\\
            0&\sqrt{1-p}
        \end{bmatrix}$ ,
     $   K_1 = \begin{bmatrix}
            0&\sqrt{p}\\
            0&0
        \end{bmatrix} $\vspace{0.2cm} and  \(0 \leq p \leq 1\).

\subsubsection{Quantum error mitigation} 
Quantum Error Mitigation (QEM) techniques attempt to reduce the effect of quantum noise on the performance of the quantum algorithm. Two QEM methods are described as follows:\\
$\noindent${\textbf{1}. \textit{Probabilistic Error Correction (PEC)}:}
It attempts to effectively mitigate the noise by inverting a well-characterized noise \cite{pec}. It works by estimating the noise in a quantum circuit and then using this information to cancel out the noise from the output of the circuit. For a noisy gate with unitary \(U'= U \circ \mathcal{N} \) first find a good approximation of \(\mathcal{N}\) and then apply its inverse before the noisy gate operation to get the noise-free gate as \(U = \mathcal{N}^{-1}U'\).\\
$\noindent${\textbf{2}. \textit{Zero noise extrapolation(ZNE)}:} It is an effective and handy technique but excessively uses the existing quantum resources for noise estimation \cite{zne1}. It works by estimating the device noise on different levels of a dimensionless parameter \(0 \leq \lambda \leq 1\), where 0 corresponds to a noiseless behaviour and 1 is a perfect match with hardware noise. The results are then extrapolated to find a noise-free behaviour of the circuit. 
 
\vspace{-0.3cm}
\subsection{Quantum machine learning}
The key advantage of Quantum machine learning (QML) algorithms is the capability to represent complex structures that are not efficiently accessible with classical computers~\cite{qml}. Parameterized Quantum Circuit (PQC) is the basis of the QML models which provide trainable circuits with the help of rotation gates. 

Quantum Neural Network (QNN) is made up of three layers input layer, trainable quantum layer and measurement layer \cite{qnn}. The input layer (or encoding layer) converts input data into the rotation angle of a predefined collection of rotation gates on qubits. The trainable quantum layer is a collection of PQCs, each having trainable parameters. Measurement layer that reads the output of each qubit and maps the class output.
\vspace{-0.3cm}
\subsubsection*{Distributed QML} 
Distributed QML can train larger models than the capacity of individual devices by circuit cutting. In circuit cutting \cite{classicalresource} a large circuit is cut at some points to create sub-circuits that can be executed on different devices. After the execution is completed the circuit is reconstructed to get the results. It consists of three phases: circuit cutting, execution and circuit reconstruction\cite{cutqc}. Cutting a quantum circuit into sub-circuits with fewer qubit requirements and after executing it on a quantum computer the reconstruction is performed. Reconstruction is done by creating a merged probability distribution for the original quantum circuit by performing repeated measurements \cite{circuittomo}. Circuit partitioning also increases the fidelity by providing an inherited noise mitigation in smaller circuits. The distributed QNN (DQNN) \cite{dqnn} can also be created using partitioned feature encoding. 

\begin{figure*}[h]
    \centering
    \includegraphics[scale=0.5]{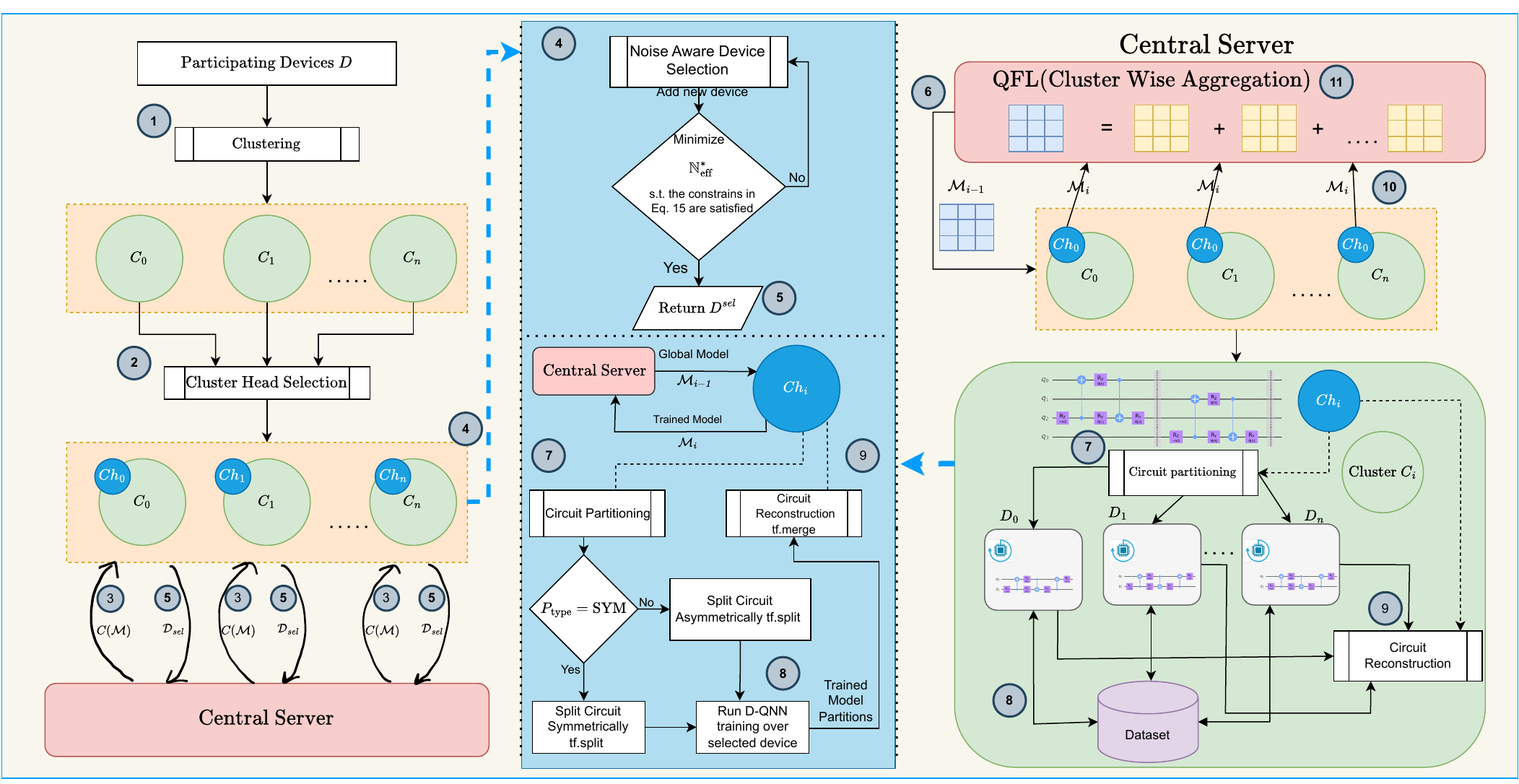}
     \caption{\small Workflow for NAC-QFL system. Steps \protect\circled{1} and \protect\circled{2} are used for cluster creation and cluster head selection. Steps \protect\circled{3}-\protect\circled{5} are used for noise-aware device selection phase. Step \protect\circled{6} initiates model distribution to the cluster heads. Steps \protect\circled{7}-\protect\circled{9} are for Distributed model training within a cluster Step \protect\circled{10} is used for trained model upload to the central server and Step \protect\circled{11} used for cluster wise aggregation.}
    \label{fig:fig3}
\end{figure*}
\vspace{-0.3cm}
\subsection{Quantum federated learning}

Federated Learning (FL) \cite{googlefed} was initially proposed for training an optimized central model by utilizing multi-node collaborative training which should be communication optimized and preserve data privacy. Localized training is performed on the edge devices over their exclusive data and a central server aggregates the global model. Further, it was extended in different IoT applications where constraint devices \cite{iotfl} are participating in FL with the goal of maximum revenues. The concept is further extended to become more inclusive by Yang et. al \cite{fedsurvey} \textit{as a general concept for all privacy-preserving decentralized collaborative machine learning techniques}. Using this definition, QFL is designed for collaborative training of QML models using quantum computers in a distributed manner. Even though the QFL is a relatively new concept, it provides many advantages such as collaborative training using a heterogeneous set of devices, data privacy and communication efficiency. The work done in the area of QFL shows that QFL is gaining significant attention in the research community \cite{qflsurvey1}. Recent works are majorly related to resource optimization, communication efficiency, network security and framework design \cite{fedqdata}\cite{qfed}. Initial proof of concept works on QFL \cite{qfl}\cite{qfed} shows the feasibility and advantages. The authors in \cite{noniid} proved that a non-IID problem exists in QFL and provided a solution with better performance.
Authors in \cite{resource} have proposed an FT-QNN which consists of an edge QNN and cloud QNN with quantum teleportation to provide resource optimization in wireless communication. The authors in \cite{FedQcomm} suggested that a ring topology for QFL is better than hub and spoke topology. Privacy issues are handled in \cite{devsecureiclr} with post-quantum cryptography secure QFL and \cite{qfldiff} using differential privacy.
\vspace{-0.3cm}
\subsection{Motivation and contributions}
This work is motivated by the limitations highlighted in previous subsections. The first limitation we observed is that quantum devices are noisy and have limited capacity~\cite{qcbench}. It is challenging to execute large-scale QML models on such noisy and limited-capacity quantum devices~\cite{dqnn}. Furthermore, QEC requires a large number of redundant qubits which is not available in NISQ-era devices. Existing QEM techniques aim to reduce the impact of quantum noise. However, these techniques only attempt to produce accurate expectation values for observable~\cite{pec} and require high computational resources growing with the circuit size~\cite{zne1}. The next limitation involves running QML on quantum devices requiring encoded datasets. Communication of such encoded datasets is costly and increases errors in existing quantum communication devices~\cite{entanglementassistedchannel}. Finally, existing quantum federated learning approaches do not consider the noise in quantum devices~\cite{qflsurvey1}. \\ 
$\noindent${\textbf{Contributions}:} The paper introduces NAC-QFL: \textbf{N}oise \textbf{A}ware \textbf{C}lustered \textbf{Q}uantum \textbf{F}ederated \textbf{L}earning system. To the best of our knowledge, this is the first work to simultaneously address noise mitigation, limited quantum device capacity, and high quantum communication costs.  Along with this, the major contributions of this work are as follows: The first contribution involves clustering the quantum devices using the K-means algorithm, utilizing the entanglement-assisted channel capacity as the distance metric. The next contribution presents noise modelling methods and a procedure for noise-aware device selection. Furthermore, we formulate a noise-aware device selection problem and propose an algorithm to solve it. Additionally, we introduce a noisy dataset for QML to demonstrate the impact of noise on the proposed system accuracy. We demonstrate that the problem is NP-complete and provide a polynomial solution.

\section{NAC-QFL system}
\label{sec:3}
This section first provides the NAC-QFL system description, followed by the workflow and methodology containing device clustering, noise-aware device selection, and model training.  
\vspace{-0.3cm}
\subsection{NAC-QFL system description and workflow} 
The NAC-QFL system consists of a QML model $\mathcal{M}$ with quantum circuit $\mathcal{Q}$, that will be trained in the federated setup of a set of $m$ quantum devices $\mathcal{D}$ = $\{D_{1}, D_{2},........D_{m}\}$. The devices are partitioned into a set of $n$ 
 clusters  $\mathcal{C} = \{C_{1},C_{2},........C_{n}\}$ based on specified similarity. The federated training is managed by the central aggregation server $\mathcal{S}$ with responsibilities of global model distribution and aggregation. A cluster $C_i$  has a cluster head $\mathcal{C}_{h_{i}}$, which is responsible for the device selection and distributed model training within the cluster.\\
Figure~\ref{fig:fig3} illustrates the NAC-QFL workflow, which shows the clustering process (step-\circled{1}), cluster head selection (step-\circled{2}), noise aware device selection (step-\circled{3}-\circled{5}), model download ($\mathcal{S}\rightarrow\mathcal{C}_{h_{i}}$)(step-\circled{6}), model training within each cluster (step-\circled{7}-\circled{9}), model upload ($\mathcal{C}_{h_{i}}\rightarrow\mathcal{S}$)(step-\circled{10}) and cluster wise aggregation (step-\circled{11}). The workflow can broadly be divided into three phases as i) Cluster formation (\S 3.2), ii) Noise-aware device selection (\S 3.3), and iii) NAC-QFL training (\S 3.3). 
\vspace{-0.3cm}
\subsection{Clustering in NAC-QFL system.}  
The NAC-QFL system uses intra-cluster communication among the devices of a cluster and the communication from the cluster head to the central server. The aim of clustering is to group the devices such that the communication cost is minimized. 
Let \(C_e\) is the entanglement-assisted channel capacity \cite{entanglementassistedchannel}  and \(d_k\) is the communication distance between two devices. The distance metric to used group devices in a cluster is given by a function $\delta_k = f(d_k,C_e)$. The NAC-QFL system uses the K-means clustering algorithm with distance metric \(\delta_k\) to form the clusters. Each cluster has a cluster head equipped with classical computing resources, denoted as \(\mathcal{R}\), required in circuit reconstruction. 
The cluster head selection metric \(\Xi\) in NAC-QFL system is calculated as,

\begin{equation}
    \Xi =  \lambda*C_e+ (1-\lambda)* R
    \label{eq:eqnew}
\end{equation}
where \( 0 < \lambda < 1\). A cluster head is elected with max \(\Xi\) among all devices. Procedure $1$ illustrates the above steps to form the clusters in the NAC-QFL system. The input and the output of the procedure are the lists of devices with the number of clusters and the cluster set with cluster heads, respectively. 
\SetAlFnt{\footnotesize}
\begin{procedure}[h]    
\caption{() \textbf{1: {Clustering in NAC-QFL system}}}
\KwIn{List of devices \(\mathcal{D} =\{D_{1}, D_{2},........D_{m}\}\), Number of clusters $n$}
\KwOut{Cluster set \(\mathcal{C} = \{C_{1},C_{2},........C_{n}\}\)}
\SetKwRepeat{Do}{do}{while}
\SetKwProg{Fn}{Function}{}{}
    \textbf{Initialize}: Randomly select $n$ devices as initial cluster centroids\;
    \Do{Flag=True or Max Iteration not completed}{
        \For{each \(\mathcal{D}_{i} \in \mathcal{D} \)}{
            calculate $\delta_k$ to each  centroid of \(\mathcal{C}_{j}\) \;
            \nonl         /*$\delta_k = f(d_k,C_e)$*/ \\
            Add \(\mathcal{D}_{i}\) to \(\mathcal{C}_{j}\) with min $\delta_k$ \;
        }
        \nonl /*Update Cluster Centroids*/\\ 
        \For{each cluster}{
            Update cluster centroid based on new mean\;
            set Flag = True \nonl/*flag indicates any update in centroid*/
        }
    }
    \nonl /*\textit{Cluster-head Selection}*/\\
    \For{each \({C}_{i} \in \mathcal{C} \)}{
        \For{each \({D}_{i} \in C_i \)}{ 
            calculate $\Xi$ using Eq.~\ref{eq:eqnew} }  
             $\mathcal{C}_{h_{i}}$$\leftarrow $Device with max $\Xi$ \; 
        }
    return \(\mathcal{C}\)\;
\end{procedure}
\vspace{-0.2cm}
\subsection{Selection of devices within a cluster}
This subsection initially presents the noise modeling methods utilized by NAC-QFL during quantum device selection. Next, it presents the procedure for noise-aware device selection. Finally, it demonstrates that this procedure can be executed in polynomial time.

\subsubsection{Noise modelling for quantum device}  
Quantum device calibration is utilized to characterize coherent noise, wherein gates are calibrated and noise levels are estimated. Table~\ref{Tab:1} illustrates different parameters related to gate error and qubits properties. The NAC-QFL system utilizes these parameters for noise modelling in devices, as they are updated after each calibration cycle. This noise model offers estimates of aggregated noise based on calibration parameters, aiding in approximating noise effects during circuit execution. 
\begin{table}[htbp]
    \centering
    \footnotesize
    \caption{\small Calibration parameters for noise model. }
    \vspace{-0.3cm}
    \label{Tab:1}
    \begin{tabular}{p{2.2cm}|p{1.1cm}|p{4.2cm}}
        \hline
        \textbf{Parameter} & \textbf{Symbol}& \textbf{Description} \\
        \hline
        Thermal relaxation time & \(T_{1}\) time& Time to loose excited state(\(\ket{1} \mapsto \ket{0}\))  \\
        \hline
        Decoherence time &\(T_{2}\) time & Time in which qubit losses . \\
        \hline
        Single-qubit gate error& \(\xi(g^1)\) & Errors due to faulty Pauli-X,Y,Z and H gates. \\
        \hline
        Two-qubit gate error& \(\xi(g^2)\)  &  Errors due to faulty CNOT gate  \\
        \hline
        Readout error & \(\xi(\pi)\) &Inaccuracies in the state measurement. \\
        \hline
        State preparation error & \(\xi(\mathbf{s}^{p}_{0/1 \rightarrow 1/0})\) & Probability of getting a state other than prepared such as measuring $\ket{1}$ when $\ket{0}$ \\
        \hline   
    \end{tabular}
    \label{tab:calibration_parameters}
\end{table}\\
\noindent Let $T_{1j}$ gives the relaxation time of $j^\text{th}$ qubit and $T_{2j}$ gives the dephasing time of $j^\text{th}$ qubit. The $\mathcal{T}_\text{eff}$ give the aggregated value of thermal relaxation time and decoherence time, \textit{i.e.,}  
\begin{equation}
\label{eq:teff}
    \mathcal{T}_\text{eff} =  \sum_{j}^{m}T_{1_j} + \sum_{j}^{m}T_{2_j}.
 \end{equation}
 The effective single qubit error is calculated as the sum of all single qubits across all qubits, \textit{i.e.,} 
\begin{equation}  
\label{eq:g1}
    \mathcal{G}^{1}_{\text{err}} =  \sum_{j}^{m}\sum_{i \in \mathbf{g}^1}  \xi(g^{i}_j),
 \end{equation}
where $\mathbf{g}^1$ is the set of all available single qubit gates and $\xi(g^{i}_j)$ is the normalized error. 
Two-qubit gates are more noisy than single-qubit gates and play a key role in deciding state fidelity.
The effective two-qubit error, denoted as $\mathcal{G}^{2}_{\text{err}}$, is calculated as the sum of all two qubits across all qubits, \textit{i.e.,}   
\begin{equation}  
\label{eq:g2}
    \mathcal{G}^{2}_{\text{err}} = \sum_{k}^{m} \sum_{j}^{m}\sum_{i \in \mathbf{g}^2} \xi(g^{i}_{jk}),
 \end{equation}
where $\mathbf{g}^2$ is the set of all available two qubit gates and $\xi(g^{i}_{jk})$ is the normalized error. The effective measurement error, denoted by $\Pi_{err}$, is the sum of measurement error for all qubits. Therefore, 
\begin{equation}
\label{eq:merr}
    \Pi_{err} = \sum_{j}^{m}\xi(\pi_j),
\end{equation}
where $\pi_j$ is the measurement error of the $j^\text{th}$ qubit. The effective state preparation error is the sum of state preparation error, \(\xi(\mathbf{s}^{p}_{0/1})\) and \(\xi(\mathbf{s}^{p}_{1/0})\), for all qubits, \textit{i.e.,} 
\begin{equation}
\label{eq:sperr}
    {S}^{p}_{\text{err}} = \sum_{j}^{m} \xi(\mathbf{s}^{p}_{0 \rightarrow 1}) + \sum_{j}^{m} \xi(\mathbf{s}^{p}_{1 \rightarrow 0}),
\end{equation}
where $\xi(\mathbf{s}^{p}_{1 \rightarrow 0})$ and $\xi(\mathbf{s}^{p}_{1 \rightarrow 0})$ state preparation error for prepare 0 measure 1 and prepare 1 measure 0 respectively. \\

\noindent NAC-QFL system considers the above parameters for estimating the effective noise model, denoted as $\mathbb{N}_{\text{eff}}$, of a quantum device \(D_{i} \in \mathcal{D}\) with $m$ number of qubits, therefore
\begin{equation}   
    \mathbb{N}_{\text{eff}} = \omega_1 \cdot \mathcal{T}_\text{eff} + \omega_2\cdot \mathcal{G}^{1}_{\text{err}}+\omega_3 \cdot \mathcal{G}^{2}_{\text{err}}+ \omega_4 \cdot \Pi_{err}+ \omega_5 \cdot {S}^{p}_{\text{err}}, 
    \label{eq:eq10}
 \end{equation}
where $\omega_1,\cdots,\omega_5$ are weights given to $\mathcal{T}_\text{eff}$, $\mathcal{G}^{1}_{\text{err}}$, $\mathcal{G}^{2}_{\text{err}}$, $\Pi_{err}$, and ${S}^{p}_{\text{err}}$ parameters, respectively. 

\subsubsection{Noise aware device selection problem}
The NAC-QFL system forms the cluster set \(\mathcal{C} = \{C_{1}, C_{2},........C_{n}\}\) using Procedure 1 and estimates the noise of each device of a given cluster using the proposed effective noise model $\mathbb{N}_{\text{eff}}$. 
Let a cluster $C_j$ consist of the device set $\mathcal{D}_j$, where $1 \leq j \leq n$. 
The estimated noise of the devices of cluster $C_j$ are denoted as $\mathbb{N}_{\text{eff}}(D_k)$, where $D_k \in \mathcal{D}_j$. Next, the NAC-QFL system arranges each device of the cluster $C_j$ in increasing order of the noise to select the least noisy devices. 
The selected devices are the least noisy and also satisfy the capacity constraint of the quantum circuit. The set of $k$ selected devices of $\mathcal{D}_j$ be denoted as $\mathcal{D}^{sel}_j$, where $\mathcal{D}^{sel}_j=\{D^{sel}_{j1},\cdots,D^{sel}_{jk}\}$ and $k=|\mathcal{D}^{sel}_j|$. 
Let \(\mathbb{N}_\text{eff}^{*}\) represent aggregated effective noise of selected devices of $C_j$ cluster,  $\Delta$ is the noise threshold, $\mathscr{P}_{lim}$ is data parallelization limit, $d$ is the dimension of a data-point in the dataset. Let $\delta_i$ be the weight associated with each device and $|D|$ is the device selection limit. The capacity of a selected device $D^{sel}_{ji}$ of cluster $C_j$ and the required capacity for the given QML model $\mathcal{M}$ are denoted as $C(D^{sel}_{ji})$ and $C(\mathcal{M})$, respectively. 
The noise-aware device selection problem can be formulated as follows: 
\begin{align}\nonumber
\noindent Minimize \hspace{0.2cm}& \mathbb{N}_\text{eff}^{*}(\mathcal{D}_j,\mathcal{M}, \Delta, \mathscr{P}_{lim}),\\ \nonumber
 \text{s.t. } 
  &  C(\mathcal{M}) \leq \sum_{i=1}^{k} C(D^{sel}_{ji}),\\ \nonumber
    & \sum_{i=1}^{k}  \delta_i*\mathbb{N}_{\text{eff}}(D^{sel}_{ji})  \leq \Delta, \\ \nonumber
    & \frac{d}{|D^{sel}_j|}\leq \mathscr{P}_{lim}, \\
    & |D^{sel}_j| \leq |D|. 
\label{eq:optim}
\end{align}
$\noindent$ Constraints of Eq.~\ref{eq:optim} illustrate as follows: The initial constraint ensures that the combined capacity of selected devices meets the required capacity of the QML model. Subsequently, a constraint prohibits the selection of devices with noise levels exceeding a specified threshold. Another constraint restricts the selection of devices with data exceeding a certain threshold. Lastly, constraint stipulates that the number of selected devices must not exceed a predefined threshold. Procedure $2$ illustrates the solution of the above noise-aware device selection problem. The input to the procedure is a cluster $\mathcal{C}_j$ with $\mathcal{D}_j$, the QML model $\mathcal{M}$ and the output of the is the set of the selected device $\mathcal{D}^{sel}_j$ and average quantum volume~\cite{quantumvolume} $\omega_j$. The procedure starts with sorting the devices based on their effective noise model $\mathbb{N}_\text{eff}$. The procedure adds a new device to the selected device list if satisfies the optimization criteria of Eq.~\ref{eq:optim}. Finally, it calculates the average quantum volume of the cluster and returns it to the central server. 
\SetAlFnt{\footnotesize} 
\begin{procedure}
\caption{()\textbf{ 2:{ Noise Aware Device Selection.}}}
\DontPrintSemicolon
\KwIn{QML model $\mathcal{M}$ \& with quantum circuit $\mathcal{Q}$, a cluster $C_j$ \; } 
\KwOut{list of selected devices $\mathcal{D}^{sel}_j$, Average Quantum Volume $w_j$ ;}
\textbf{Initialize}: {$\mathcal{D}^{sel}_j$ $\gets$ 0} is initialized as an empty list\;
$\mathcal{D}_j^{'}$ $\gets$ sort($\mathcal{D}$); sort device list based on their noise model\;
\For{each device $D_i\in \mathcal{D}_j^{'}$}
{
\uIf{
Adding \(D_i\) minimizes $\mathbb{N}_\text{eff}^{*}$\;
under constraints given Eq.~15}
{add $D_i \gets \mathcal{D}_{sel}$\;}
}
return $D_{sel}$\;
\nonl /*average quantum volume calculation*/\\
$w_j\gets \sum_{d_i \in \mathcal{D}^{sel}_j} w(d_i)$ \;
\nonl\Return Selected Device List $\mathcal{D}^{sel}_j$, Average Quantum Volume $\omega$  \;
\end{procedure}
\vspace{-0.5cm}
\begin{theorem}
The noise-aware device selection problem from a given pool of quantum devices \(\mathcal{D}\) to execute a circuit of required capacity is an NP-complete problem.  
\end{theorem}
\begin{proof}
The noise-aware device selection problem can be mapped to the subset sum problem, which is known to be NP-complete. The subset sum problem states that to identify a subset of positive integers from a given set whose sum falls within a specified range. Similarly, the noise-aware device selection problem states that find a subset \(\mathcal{D}^{sel}_j \subseteq \mathcal{D}_j\) such that $C(\mathcal{M}) \leq \sum_{i=1}^{k} C(D^{sel}_{ji})$ with other given constraints. 
\end{proof}
\begin{corollary}
The noise-aware device selection problem with given data parallelization constant \(p\) will become a polynomial problem with \(O(n^p)\).
\end{corollary}
\begin{proof}
In the subset sum optimization problem, where the subset sizes are fixed, the problem can be solved in polynomial time due to the reduced search space. Similarly, in the device selection problem with a parallelization limit, the number of selected devices is fixed by using the constraints ($\frac{d}{|D^{sel}_j|}\leq \mathscr{P}_{lim}$ and $|D^{sel}_j| \leq |D|$). The device selection problem can be solved with \(O(n^p)\) polynomial complexity, where $p=\min\{p',|D|\}$ and $p'$ be the number of selected devices that satisfy $\frac{d}{|D^{sel}_j|}\leq \mathscr{P}_{lim}$.  
 \end{proof}
\subsection{NAC-QFL system to train QML model}
The cluster head of each cluster in the NAC-QFL system intimates about the selected devices to the central server. 
The central server then initiates the training of the initial global QML model by distributing the model to the clusters. Each cluster trains the given model using local data on the selected devices and returns it to the central server. The central server aggregates the trained model and distributes it to the clusters. The model training process repeats till the convergence. The complete steps to train the QML model using the NAC-QFL system are illustrated in Algorithm~\ref{algo1}. The input to the algorithm is a set of quantum devices which are grouped into\(\mathcal{C}\) clusters, initial QML model $\mathcal{M}_{0}$, and accuracy threshold and the output is the trained QML model $\mathcal{M}_{f}$.  
Each cluster head locally trains the model in a distributed setup on the selected devices using Procedure 3. After the distributed training is completed the cluster head shares the trained model to the central server and the central server aggregates the models received from all cluster heads. This process is repeated by sharing the updated model with cluster heads till the termination criteria are satisfied. 
\vspace{-0.2cm}
\SetAlFnt{\footnotesize}
\begin{algorithm}[htbp]
\footnotesize
\caption{\textbf{NAC-QFL: Quantum Federated learning with clustering.}}
\DontPrintSemicolon
\label{algo1}
\KwIn{Set of quantum devices $\mathcal{D}=\{D_1, \cdots, D_m\}$, grouped into cluster set $\mathcal{C} = \{C_{1}, \cdots, C_{n} \}$, QML model $\mathcal{M}$ with quantum circuit $\mathcal{Q}$, threshold \textit{Th}, maximum communication rounds $\kappa$\; } 
\KwOut{Trained model $\mathcal{M}_{f}$, on selected participants using \textbf{NAC-QFL} ;}
\Do{($\mathcal{M}_{acc} \le Th$ \& i $\leq$ $\kappa$)}{
Centralize Server $\mathcal{S}$ shares the $\mathcal{M}_i$ to all cluster head $\mathcal{C}_{h_i}$\;
\For{each cluster $C_k \in \mathcal{C}$}{
\nonl /*\textit{Participants selection for each cluster}*/\\
Call $\textbf{Procedure 2}$ for participants selection.\;
\nonl /*\textit{Procedure for circuit cutting}*/\\
Call $\textbf{Procedure 3}$  for circuit cutting in $p$ partitions \;
\nonl /*\textit{Distributed Model Training on selected device}*/\\
     \For{ each device in  $\mathcal{D}^{sel}_j$}
      {Train assigned model partition over shared local dataset \;}
      \nonl /*\textit{Circuit Reconstruction}*/\\
      Perform circuit reconstruction at each cluster head \;
\nonl /*\textit{ Model Upload}*/\\ 
Upload the trained model $\mathcal{M}_i$ to the central server $\mathcal{S}$
}
\For{each cluster $C_i\in C$}{ \nonl /*\textit{Model Aggregation at central server}*/\\
Obtain the model from each cluster aggregate using Eq.~\eqref{eq:fedavg}\;
Send the Updated model $\mathcal{M}_{i+1}$ to all the clusters.\;
Test the performance of $\mathcal{M}_{i+1}$ on the testing dataset. \;
}
}
\nonl\Return Trained Aggregated model $\mathcal{M}_{f}$ \;
\end{algorithm}
\vspace*{-0.4cm}
\subsubsection{Distributed QFL training} It consists of three key tasks: i). Circuit partitioning ii). Circuit reconstruction and iii). Model upload. For distributed training the quantum circuit, is partitioned into sub-circuits using Procedure $3$ that can be trained on the shared dataset within the cluster. Procedure $3$ requires a QML model, selected device list and partition type as input and provides the sub-circuit list as output. It begins by checking if the selected device's aggregated capacity can execute the given circuit otherwise it will generate an error. After that based on partition type, it partitions the given circuit and returns the sub-circuit. After that sub-circuits are merged to reconstruct the quantum circuit and then the trained model is uploaded to the central server for aggregation. 
\SetAlFnt{\footnotesize}
\begin{procedure}
\caption{() \textbf{3:{ Circuit partitioning}}}  
\state \KwIn{QML model $\mathcal{M}$ (corresponding circuit $\mathcal{Q}$), list of selected devices $\mathcal{D}_{sel}$, partition type $\mathcal{P}_{\text{type}}$} 
\KwOut{list of partitions $\mathcal{P}$ ;}
\textbf{Initialize}: {$\mathcal{P}$ $\gets$ 0},  $\mathcal{P}$ is initialized as an empty list. \; 
\nonl /* \textit{C(.)} returns the qubit capacity of a circuit or device */\;
\uIf{ (C($\mathcal{Q}$) $\leq$ (\(\sum\) C($\mathcal{D}_{sel}$)))}
{
\uIf{($\mathcal{P}_{\text{type}}$ $\gets$ $\mathrm{SYM}$)}
{ Perform symmetric partitioning to create partitions $p_i$ of $\mathcal{Q}$, $\mathcal{P}.append(p_i$) \;}
\uElse
{ Perform asymmetric partitioning to create partitions $\mathcal{Q}$, $\mathcal{P}.append(p_i$) \;}
}
\uElse
{
Partitioning error \;
\nonl /* Aggregated device capacity is not sufficient to execute the model */\;
}

\nonl\Return Partition List $\mathcal{P}$ \;
\end{procedure}
\subsubsection{Model aggregation}
After sharing the global model the central server waits till it receives the update from each cluster head.
After each communication round of NAC-QFL the trained models received by the central server are aggregated using the quantum version of FedAvg as given in Eq.~\eqref{eq:fedavg}). Quantum volume\cite{quantumvolume} is a single number metric calculated by the device performance over the random circuits. Each cluster is assigned a weight $w$ which is calculated as the average quantum volume of selected devices.  
\begin{equation}
\label{eq:fedavg}
    \Theta_{i+1}^{S} = \sum_{k \epsilon K} w_{k}.\Theta_{i}^{k},
    \end{equation}
where $\Theta$ represents the trained parameters which are averaged based on weights $w$ assigned to every cluster.
\section{Performance evaluation}
\label{sec:4} 
This section first describes the datasets used, the QML model, the experimental setup and the evaluation criteria. After that, it provides results and discussion for the NAC-QFL under different settings.
\vspace*{-0.2cm}
\begin{table*}[htbp]
\centering
\small
\caption{\label{tab:2} \small Average classification results (Accuracy for MNIST-Bin, MNIST-Full, P-MNIST, MNIST-Bin(N), P-MNIST(N) for testing. NA-Noise Aware Device Selection, R-Random Device Selection; N-Noisy dataset with noisy label.}
\vspace{-0.4cm}
\begin{tabular}{cccccccccc}
\toprule
\textbf{Setting} & \textbf{Method} & \multicolumn{1}{c}{\textbf{MNIST-Bin}} &\multicolumn{1}{c}{\textbf{MNIST-Full}}&
\multicolumn{1}{c}{\textbf{P-MNIST}}&
\multicolumn{1}{c}{\textbf{MNIST-Bin(N)}} &
\multicolumn{1}{c}{\textbf{MNIST-Full(N)}} & 
\multicolumn{1}{c}{\textbf{P-MNIST(N)}} \\
\midrule
\multirow{ 3}{*}{\makecell{\textbf{S1}}} 
&QNN & 56.28 $\pm$ 0.45 & 48.86 $\pm$ 4.32 & 48.66 $\pm$ 3.47 & 41.24 $\pm$ 4.96& 39.29 $\pm$ 4.72 & 40.29 $\pm$ 1.53 \\  
&QNN(PEC) & 78.46 $\pm$ 5.10 &  62.65 $\pm$ 5.64 & 52.50 $\pm$ 3.90 & 54.77 $\pm$ 2.98&  48.03 $\pm$ 0.88 & 48.64 $\pm$ 0.46 \\  
&QNN(ZNE) &82.35 $\pm$ 3.10 &  64.00 $\pm$ 3.64 & 54.95 $\pm$ 2.55 & 56.26 $\pm$ 1.2&  51.82 $\pm$ 0.81 & 52.97 $\pm$ 0.67 \\  
\midrule
\multirow{4}{*}{\makecell{\textbf{S2}}}
&DQNN(R) & 84.01 $\pm$ 4.20 &  85.65 $\pm$ 2.12 & 82.70 $\pm$ 3.10 & 80.34 $\pm$ 4.13&  74.73 $\pm$ 3.76 & 70.53 $\pm$ 3.64 \\  
&DQNN(NA)& 84.45 $\pm$ 2.60 &  85.90 $\pm$ 2.43 & 84.25 $\pm$ 1.30 & 83.85 $\pm$ 3.82&  75.81 $\pm$ 3.71 & 72.10 $\pm$ 2.49 \\   
&DQNN(R+ZNE) & 89.22 $\pm$ 4.80 &  90.50 $\pm$ 1.94 & 86.06 $\pm$ 1.10 & 85.38 $\pm$ 1.91&  78.43 $\pm$ 1.63 & 75.50 $\pm$ 2.28 \\  
&DQNN(NA+ZNE) & 88.10 $\pm$ 3.20 &  87.22 $\pm$ 1.39 & 90.45 $\pm$ 0.30 & 85.29 $\pm$ 0.58&  82.75 $\pm$ 0.84 & 84.97 $\pm$ 1.19 \\ 
\midrule
\multirow{4}{*}{\makecell{\textbf{S3}}}
&DQNN(R) & 82.33 $\pm$ 3.60 &  83.10 $\pm$ 3.64 & 81.50 $\pm$ 4.40 & 81.92 $\pm$ 4.73&  78.36 $\pm$ 3.65 & 76.50 $\pm$ 2.84 \\   
&DQNN(NA) & 84.88 $\pm$ 3.50 &  83.29 $\pm$ 2.76 & 83.45 $\pm$ 3.18 & 81.67 $\pm$ 3.40&  77.97 $\pm$ 2.48 & 78.93 $\pm$ 1.39 \\  
&DQNN(R+ZNE) & 88.10 $\pm$ 2.30 &  85.30 $\pm$ 2.08 & 84.43 $\pm$ 1.73 & 84.01 $\pm$ 1.58&  80.50 $\pm$ 1.66 & 80.58 $\pm$ 1.31 \\  
&DQNN(NA+ZNE) & 88.02 $\pm$ 3.40 &  87.46 $\pm$ 0.98 & 90.15 $\pm$ 0.66 & 85.39 $\pm$ 0.32&  81.25 $\pm$ 0.64 & 82.67 $\pm$ 0.86 \\  
\midrule
\multirow{ 4}{*}{\makecell{\textbf{S4}}}
& FedAvg(R) & 90.83 $\pm$ 0.45  & 90.57 $\pm$ 2.32 & 88.65 $\pm$ 1.48 & 90.46 $\pm$ 1.47& 77.33 $\pm$ 0.48 & 84.25 $\pm$ 1.85 \\   
&FedAvg(NA) & 94.75 $\pm$ 0.72 & 90.83 $\pm$ 1.34  & 88.31 $\pm$ 1.93 & 92.58 $\pm$ 0.26 &  79.64 $\pm$ 0.06& 86.85 $\pm$ 1.57 \\  
&FedAvg(NA+PEC) & 96.32 $\pm$ 0.16 & 93.41 $\pm$ 0.58 &91.27 $\pm$ 1.42 & 94.29 $\pm$ 0.79 & 81.49 $\pm$ 0.96 & 88.36 $\pm$ 0.96 \\  
&FedAvg(NA+ZNE) & 97.53 $\pm$ 0.82 & 94.82 $\pm$ 0.40 & 91.73 $\pm$ 0.71 & 94.65 $\pm$ 0.38& 83.71 $\pm$ 0.11 & 90.49 $\pm$ 0.28 \\
\midrule
\multirow{4}{*}{\makecell{\textbf{S5}}} 
& FedAvg(R) & 92.52 $\pm$ 0.89 & 91.64 $\pm$ 2.85 & 89.45 $\pm$ 1.37 & 90.03 $\pm$ 1.63& 89.15 $\pm$ 1.85 & 86.41 $\pm$ 2.59 \\  
&FedAvg(NA) & 96.03 $\pm$ 0.91 & 92.42 $\pm$ 1.73  &  90.95 $\pm$ 1.03& 93.27 $\pm$ 0.92 &  90.46 $\pm$ 0.24& 88.59 $\pm$ 1.36\\ 
&FedAvg(NA+PEC) & 97.94 $\pm$ 0.04 & 94.01 $\pm$ 0.58&91.15 $\pm$ 1.85 &\textbf{ 95.84$\pm$ 0.10} & 89.36 $\pm$ 0.03 & 89.26 $\pm$ 1.11 \\ 
&FedAvg(NA+ZNE) &  \textbf{98.02 $\pm$ 0.73} & \textbf{95.15 $\pm$ 0.64} & \textbf{92.19 $\pm$ 0.63} & 95.22 $\pm$ 0.16& \textbf{90.49 $\pm$ 0.27} & \textbf{90.55 $\pm$ 0.85} \\
\bottomrule
\end{tabular}
\end{table*}
\subsection{Datasets model and experimental setup}
$\noindent$ $\bullet$\textbf{Dataset:} The MNIST handwritten dataset \cite{lecun1998mnist} is used for binary classification and multi-class classification. PneumoniaMNIST (P-MNIST), a chest X-ray dataset, is also used for experiments available as a part of MedMNIST dataset\cite{medmnistv2}. We have also created a noisy version of datasets \footnote{https://sites.google.com/itbhu.ac.in/quan/home} by adding noise to the existing dataset.
\\$\noindent$ $\bullet$\textbf{QNN model:} The QNN model is used for NAC-QFL training consists of input-layer, cropping2D, average-pooling, flatten, tf.split, quantum layer(encoding, training and measurement) and tf.concat. \\
$\noindent$ $\bullet$\textbf{Software and libraries:} 
To simulate the quantum device, we have used the Qiskit \cite{Qiskit} library which provides support for a variety of fake instances with noise models from \textit{qiskit\_ibm\_runtime. fake\_provider}. The DQNN is implemented using Pennylane \cite{PennyLane} and Keras libraries. The quantum layers are converted to Keras layers, \textit{tf.split} is used to split the model and \textit{tf.concatenate} is used to merge the splited models. To implement the noise mitigation mitiq \cite{LaRose2022mitiqsoftware}, a python package is utilized as \textit{mitiq.zne} for ZNE and \textit{mitiq.pec} for PEC. \\
$\noindent$ $\bullet$\textbf{Hardware:} 
The NSM-ParamShivay \cite{paramshivay} supercomputer is used to implement NAC-QFL. Each fake device instance and cluster head are executed on an Intel Xeon skylake 6148 with 20-node processor at 2.30 GHz, with 192 GB allocated DDR4 memory. Each cluster is kept at a different compute node so that the communication with the central server can be simulated. The central server is connected to other nodes via 100 Gbps Mellanox infiband link which is used to simulate the noisy quantum channels. \\
$\noindent$ $\bullet$ \textbf{Experimental setup} The experiments are performed under five different settings as follows: \textbf{S1} \(\rightarrow\) Single system - IBM\_mumbai (27-Qubit backend) on which a QNN is executed with and without noise mitigation. \textbf{S2} \(\rightarrow\) Cluster \(\mathcal{C}_1\) is used for running DQNN with SYM partitioning with and without NA device selection. \textbf{S3} \(\rightarrow\) Cluster \(\mathcal{C}_1\) is used for running DQNN with ASYM partitioning with and without NA device selection. \textbf{S4} \(\rightarrow\) Three clusters (\(\mathcal{C}_1, \mathcal{C}_2, \mathcal{C}_3\)) with SYM partitioning in each cluster and Cluster QFL. \textbf{S5} \(\rightarrow\) Similar to \textbf{S4} with ASYM partitioning. The cluster configuration provided in Table~\ref{tab:clusterconf} includes the device pool and capacity. \\
$\noindent$ $\bullet$ \textbf{Evaluation strategy:} The evaluation is designed to serve the primary motive of testing NAC-QFL such as to improve the local performance and global model generalization ability in the presence of noise. To evaluate following strategies are utilized: \\
\noindent \textbf{(1) Cluster performance:} It estimates: \textit{how noise aware device selection and noise mitigation performs in cluster-wise DQNN training?}\\ 
\noindent \textbf{(2) Global performance:} It is the weighted average over cluster performance and determines: \textit{Is cluster QFL is any better than individual cluster and how it deviates from cluster performance?}\\
\noindent \textbf{(3) Noisy Dataset:} Performance is evaluated for noisy dataset:\textit{\textit{ how the proposed model performs on noisy dataset?}}\\
\noindent \textbf{(4) Noisy communication channel:} Performance is evaluated for noisy communication channel: \textit{\textit{ how does the proposed model perform when communication channels are noisy?}}
\vspace{-0.4cm}
\begin{table}[ht]
    \footnotesize
    \centering
    \caption{Cluster configuration. }
    \vspace{-0.3cm}
    \begin{tabular}{p{1.2cm}p{7cm}}
    \toprule
    \textbf{Cluster-ID} & \textbf{Devices-IBM\_fakebackend}\\ \midrule
        \(C_1\) & Oslo(5) Perth(5) Lagos(7)Melbourne(14) Mumbai(27) Hanoi(27) \\
       \(C_2\)  & Jakarta(5) Manila(5) London(5) Rueschlikon(16) Kolkata(27) \\
        \(C_3\)  & Lima(5) Narobi(5) Casablanca(7) Geneva(27) Algiers(27) Sydney(27)\\
        \hline
    \end{tabular}
    \label{tab:clusterconf}
    \vspace{-.4cm}
\end{table}
\subsection{Results and discussion} 
\begin{figure*}[htbp]
  \begin{subfigure}{0.24\textwidth}
     \centering
        \includegraphics[scale=.4]{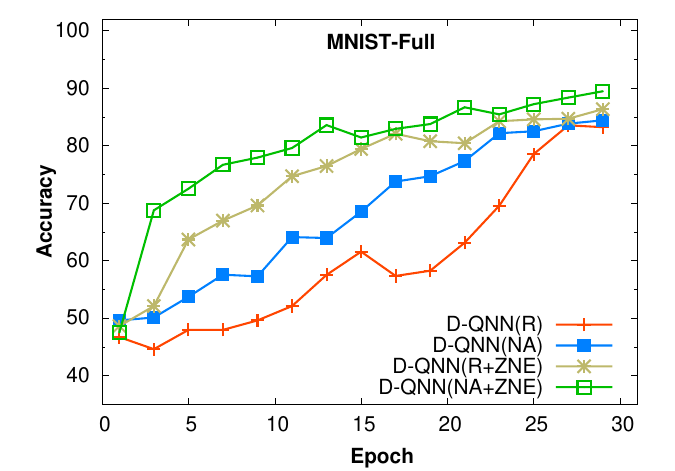}
        \vspace{-.6cm}
        \subcaption{\scriptsize Single cluster SYM-Partition} 
        \label{fig:fig5a}
  \end{subfigure}%
  \begin{subfigure}{0.24\textwidth}
     \centering
          \includegraphics[scale=.4]{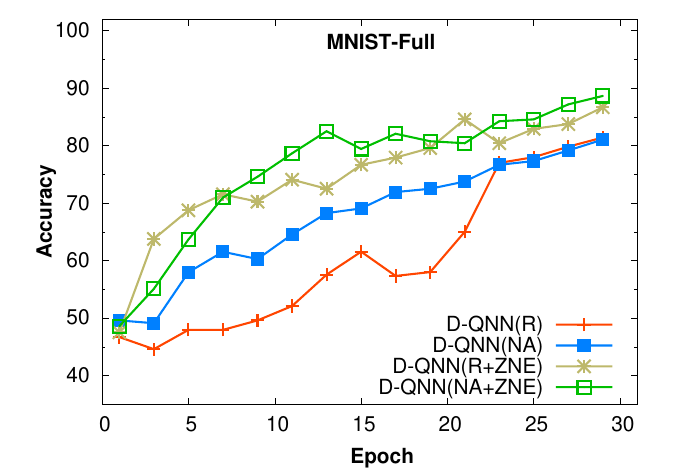}
        \vspace{-.6cm}
        \subcaption{\scriptsize Single cluster ASYM-Partition}
        \label{fig:fig5b}
  \end{subfigure}
  \begin{subfigure}{0.24\textwidth}
     \centering
         \includegraphics[scale=.4]{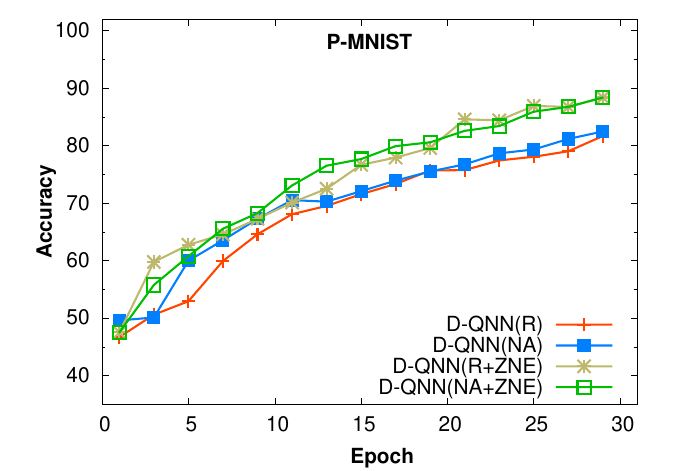}
        \vspace{-.6cm}
        \subcaption{\scriptsize Single cluster SYM-Partition}
        \label{fig:fig5c}
  \end{subfigure}
  \begin{subfigure}{0.24\textwidth}
     \centering
        \includegraphics[scale=.4]{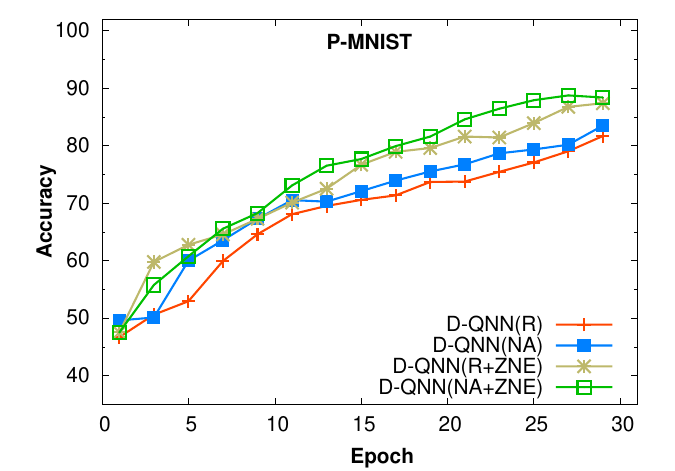} 
         \vspace{-.6cm}
        \subcaption{\scriptsize Single cluster ASYM-Partition}
        \label{fig:fig5d}
  \end{subfigure}
  \begin{subfigure}{0.24\textwidth}
         \vspace{0.5cm}
     \centering
        \includegraphics[scale=0.4]{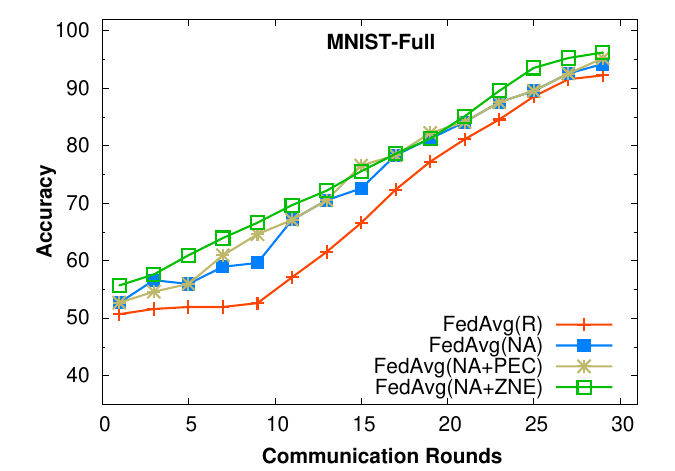}
        \vspace{-.6cm}
        \subcaption{\scriptsize FedAvg-SYM}
        \label{fig:fig5e}
  \end{subfigure}
  \begin{subfigure}{0.24\textwidth}
     \centering
        \includegraphics[scale=0.4]{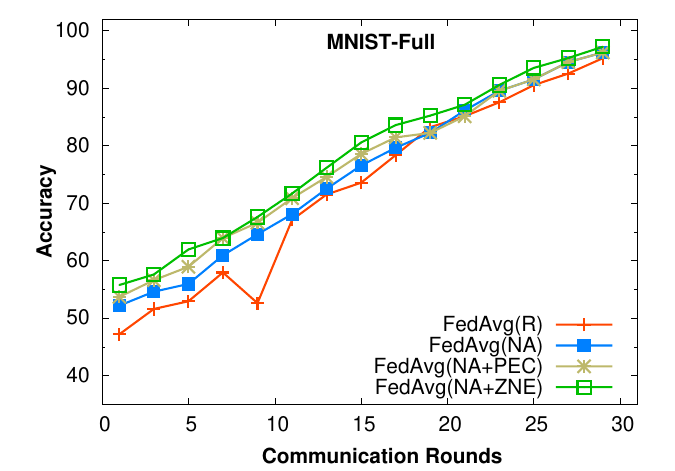}
        \vspace{-.6cm}
        \subcaption{\scriptsize FedAvg-ASYM}
        \label{fig:fig5f}
  \end{subfigure} 
  \begin{subfigure}{0.24\textwidth}
     \centering
        \includegraphics[scale=0.4]{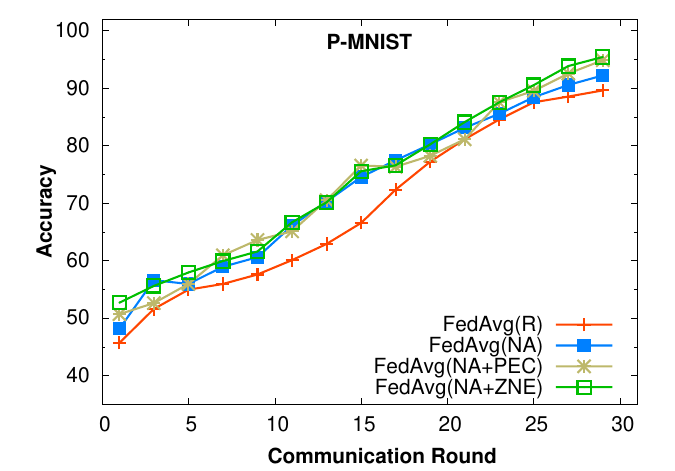}
        \vspace{-.6cm}
        \subcaption{\scriptsize FedAvg-SYM}
        \label{fig:fig5g}
  \end{subfigure}
  \begin{subfigure}{0.24\textwidth}
     \centering
        \includegraphics[scale=0.4]{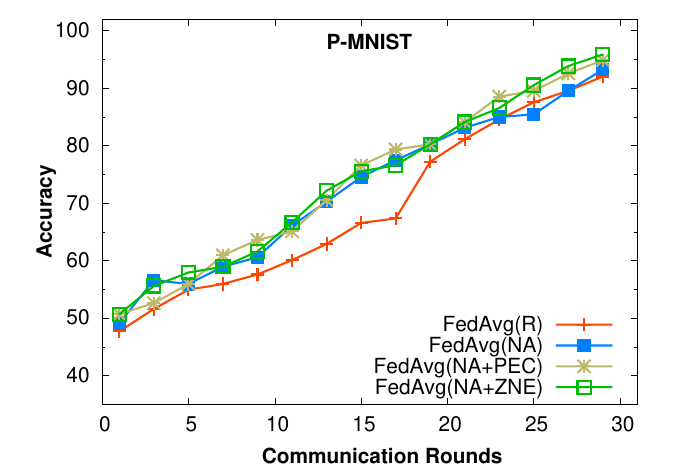}
        \vspace{-.6cm}
        \subcaption{\scriptsize FedAvg-ASYM}
           \label{fig:fig5h}
  \end{subfigure}
  \caption{\label{fig:fig5} \small Training accuracy results. Parts (a)-(d) are plots for the training accuracy for the single cluster performance. Parts (e)-(h) are average testing accuracy for the QFL with the communication rounds}
  \vspace{-.2cm}
\end{figure*}

We validate the performance of NAC-QFL by conducting a series of experiments on different datasets, including MNIST-Bin/Full, P-MNIST, MNIST-Bin/Full(N), and P-MNIST(N). Experiments are performed under two device selection strategies Random-R and Noise Aware-NA. We extend our evaluations to study the performance of NAC-QFL over both noiseless and noisy communication channels, thereby simulating real-world scenarios.\\
\noindent \textit{\textbf{4.2.1 Cluster performance:}} Initially, we performed experiments using localized DQNN training within a cluster. The results for the MNIST-Full and P-MNIST training accuracy plots are shown in Figure~\ref{fig:fig5a}- Figure~\ref{fig:fig5d}. Notably, the accuracy of DQNN(R) is recorded at 85\%, which lags behind DQNN(NA) at 90\% in both S2 and S3 settings. This is due to the fact that noise-aware selection prioritizes devices with better noise parameters which reduces the impact of system noise on model performance. It can also be seen that DQNN(NA+ZNE) yields a smoother curve and faster convergence, due to effective noise mitigation on the selected device. The testing accuracy presented in Table~\ref{tab:2} for settings S2 and S3 further elucidates the performance disparities. In the case of noise mitigation techniques, DQNN(R) achieves 85\% accuracy for MNIST-Full and 90\% for P-MNIST. While comparing the effects of noise mitigation for S1 setting, it becomes evident that these techniques are particularly efficacious in DQNN, owing to their smaller circuit size.\\
$\noindent \blacksquare$ \textbf{Observations:}
Cluster performance is evaluated to verify the advantages provided by the circuit cutting and noise-aware device selection. It is observed that the noise-aware device selection provides better accuracy. Another key observation is that there is a significant improvement in accuracy in training a model with and without noise mitigation.    
\\
\noindent \textit{\textbf{4.2.2 NAC-QFL performance:}} Subsequently, our experiments extended to the cluster-based QFL framework, utilizing settings S4 and S5. The accuracy plots are shown in Figure~\ref{fig:fig5e}-\ref{fig:fig5h}. While comparing the results of single-cluster configurations with NAC-QFL, a notable improvement in performance is observed. Also, more smooth curves are observed in the case of NAC-QFL due to the averaging effect of NAC-QFL. Furthermore, as the communication rounds increased, the higher accuracies are achieved faster as compared to single clusters. The average accuracy results are shown in Table~\ref{tab:2} which shows the best accuracy achieved in the binary classification with noise-aware device selection and asymmetric partitioning as 98.75\%. Comparing the Device selection strategy, noise-aware device selection outperforms random device allocation across different settings.  \\
\vspace{-0.4cm}
\begin{table}[htbp]
    \centering
\footnotesize
     \caption{\label{tab:tab4} \small Avg F1-score.}
     \vspace{-0.4cm}
    \begin{tabular}{p{0.8cm}p{2.3cm}p{0.9cm}p{0.8cm}p{0.9cm}p{0.8cm}}
    \toprule
      \textbf{Setting} & \textbf{Method} & \textbf{MNIST-Full}  & \textbf{P-MNIS}T & \textbf{MNIST-Full(N)}  & \textbf{P-MNIST(N)}  \\
       \midrule
        \multirow{2}{*}{\textbf{S4}}& FedAvg(NA) & 92.45& 88.01 & 76.90 & 85.90  \\
                                    &FedAvg(NA+ZNE) &94.10 & 91.20 & 82.82& 89.95  \\ \midrule
       \multirow{2}{*}{\textbf{S5}} &FedAvg(NA) & 91.92& 89.95 & 89.54 & 88.15  \\
                                     &FedAvg(NA+ZNE) & 95.05 &91.90& 90.00&  90.15 \\ \bottomrule
          
    \end{tabular}  
\end{table}

\begin{figure}[htbp]
\begin{subfigure}[b]{0.495\linewidth}
         \centering
         \includegraphics[scale=0.4]{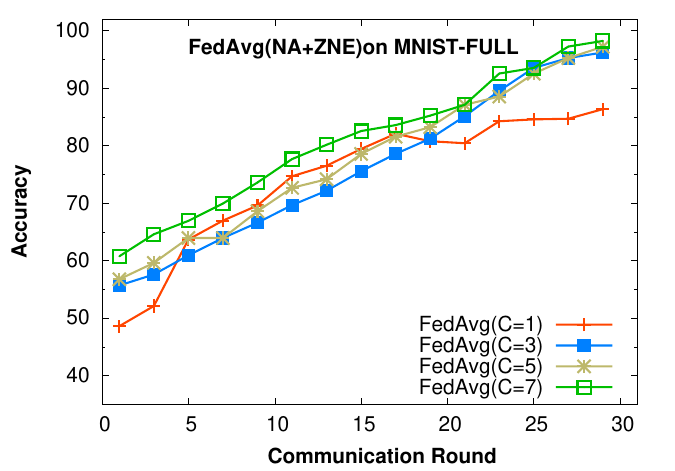}
        \subcaption{}
        \label{fig:fig6a}
    \end{subfigure}
        \begin{subfigure}[b]{0.495\linewidth}
        \centering
      \includegraphics[scale=0.4]{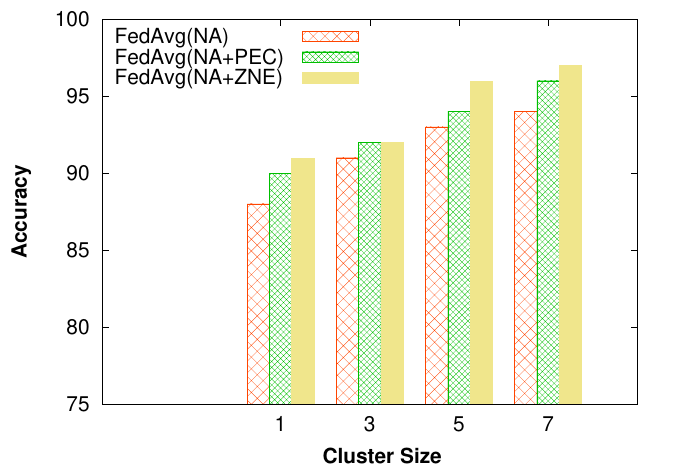}
        \subcaption{}
        \label{fig:fig6b}
    \end{subfigure}
    \vspace{-0.8cm}
    \caption{\small QFL performance with number of cluster varying from C=1,3,5,7. Part (a) shows training accuracy plot with number of clusters increased. Part (b) shows the accuracy values for different settings.}
    \label{fig:fig6}
\end{figure}

\begin{figure}[h]
    \centering
        \includegraphics[width=.70\linewidth]{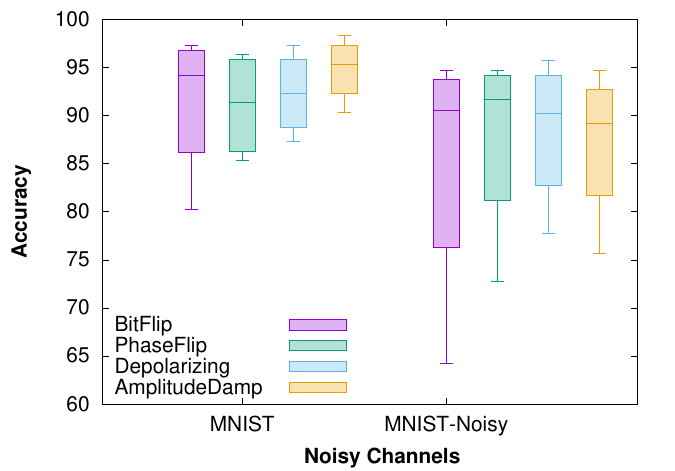}
     \caption{\small QFL performance over different noisy channels. The graph shows the variation from noise factor varying from 0.01 to 0.25\%}
     \label{fig:fig7}
     \vspace{-0.3cm}
\end{figure}

\noindent Relying solely on accuracy may overlook important aspects of model performance. So, we also evaluated the F1-score as shown in Table~\ref{tab:tab4}. For instance, in the S4-FedAvg(R), the F1-score lag behind accuracy, while improving notably in S4-FedAvg(NA). A similar trend follows in the case of S5 setting. Notable results can be seen for noisy datasets where the F1-score is significantly low in FedAvg(NA) in S4 and S5 settings but improved with noise mitigation. Also if highly noisy devices are selected the F1-score degrades as compared to noise-aware device selection. \\
$\noindent \blacksquare$ \textbf{Observations:} The NAC-QFL is evaluated to verify the performance of clustering and DQNN training over noise-aware device selection. The effect of noise-aware device selection with ASYM partitioning provides better performance due to the individual performance of devices. The FedAvg aggregation extends the same advantage to the global model performance. \\
\noindent \textit{\textbf{4.2.3 Performance with number of cluster:}} To understand the effect of varying number of clusters and NAC-QFL performance we varied the number of clusters as 1,3,5,7. The results presented in Figure~\ref{fig:fig6} clearly show that by increasing the number of clusters leads to a notable improvement in the overall accuracy of the global model. In addition, the analysis reveals that higher numbers of clusters facilitate achieving the same accuracy at a faster rate. The experiments conducted on the MNIST-Bin and MNIST-Full datasets demonstrate differing effects of the number of clusters. While the influence on MNIST-Bin is limited beyond three clusters, resulting in marginal accuracy improvements, the impact on MNIST-Full is notably significant as accuracy improves from 95\%(C=3) to 98\%(C=7). This result is due to the limited learning capacity in the case of three clusters which fails to capture the complexities of a large dataset. The results improved with five and seven clusters.\\
$\noindent \blacksquare$ \textbf{Observations:} The primary observation is that increasing the number of clusters not only increases the accuracy but also leads to faster convergence.
\\
\noindent \textit{\textbf{4.2.4 Performance over the noisy datasets:}} In the presence of noisy datasets, both cluster DQNN and QFL models exhibit degraded performance, as shown in Table~\ref{tab:2} for the noisy versions of MNIST and P-MNIST. Comparing the results with non-noisy datasets, we observe a significant decrease in model accuracies for MNIST-Full, ranging from 3\% to 10\% across different settings. In S4, for instance, the accuracy of FedAvg(NA) on average drops from 90\% to 80\%. Similar trends are shown in different settings also. The primary reason behind the performance drop is the multiplicative effect of system noise over noisy datasets. However, applying noise mitigation techniques yields promising results. Specifically, FedAvg(NA+ZNE) achieves around a 3\% performance gain across different settings, indicating a significant reduction in the impact of noise. This observation underscores the potential of noise mitigation techniques to improve accuracy, even in the presence of noisy datasets.\\
$\noindent \blacksquare$ \textbf{Observations:} The presence of noise in the dataset leads to a significant performance drop. However, this degradation is mitigated by utilizing noise-aware device selection and noise mitigation techniques.\\
\noindent \textit{\textbf{4.2.5 Noise Mitigation Analysis:}} The effectiveness of noise mitigation techniques in improving model accuracy is shown in Table~\ref{tab:2} and Figure~\ref{fig:fig5}. For S1 setting, the performance of QNN for MNIST-Full improves from 48\% to 64\% which is common for other scenarios. Notably, ZNE outperforms PEC despite being more resource-intensive. This trend persists across other settings. In Figure~\ref{fig:fig5}, we observed smoother curves across all settings where noise mitigation is used compared to scenarios without noise mitigation. Moreover, intriguing findings arise in case of noisy datasets where the performance improved by these noise mitigation techniques. The F1-score are also improved as shown in Table~\ref{tab:tab4}.
\\
$\noindent \blacksquare$ \textbf{Observations:} In comparison to QNN, noise mitigation techniques demonstrate greater effectiveness in DQNN  since their effectiveness increases on smaller circuits.
\vspace*{-0.1cm}
\begin{table}[htbp]
    \centering
\footnotesize
     \caption{\label{tab:3}\small NAC-QFL performance with noisy communication}
     \vspace{-0.3cm}
    \begin{tabular}{p{1cm}p{1.5cm}p{1.05cm}p{1.05cm}p{1.05cm}p{1.05cm}}
    \toprule
      \makecell{\textbf{Noise}\\\textbf{Channel}} &\makecell{ \textbf{DataSet} $\downarrow$\\\textbf{Intesity} $\rightarrow$} & \makecell{\\ \textbf{0.01}} &\makecell{\\ \textbf{0.05}} & \makecell{\\ \textbf{0.10}} &\makecell{\\ \textbf{0.25}} \\ \midrule
       
        \multirow{2}{*}{BitFlip}& MNIST-Bin & 97.33 $\pm$ 0.6& 96.22 $\pm$0.4 & 92.05 $\pm$ 0.8& 80.22$\pm$0.3\\
         &MNIST-Bin(N) & 94.29 $\pm$0.5 & 92.00$\pm$ 0.2& 88.22$\pm0.5$  & 68.44$\pm$ 4.4\\ 
       \multirow{2}{*}{PhaseFlip} &MNIST-Bin  &94.26 $\pm$ 0.4&  93.65 $\pm$ 0.3& 90.35 $\pm$ 3.1& 83.27 $\pm$ 4.3\\
       &MNIST-Bin(N) &92.97 $\pm$ 0.9&88.26 $\pm$ 1.4&  87.65 $\pm$ 3.3& 82.35 $\pm$ 4.1\\ 
       \multirow{2}{*}{Depolarizing} &MNIST-Bin  &96.87 $\pm$ 0.4 &  95.66 $\pm$ 0.5 & 87.37 $\pm$ 2.1 & 85.28 $\pm$ 0.3 \\
       &MNIST-Bin(N) &  94.54$\pm$ 0.9&93.68 $\pm$ 0.7 &  89.60 $\pm$ 0.3 & 72.85 $\pm$ 3.1\\ 
       \multirow{2}{*}{\makecell{Amplitude\\Damping}}& MNIST-Bin &97.45 $\pm$ 1.1 &  94.74 $\pm$ 0.4 & 90.68 $\pm$ 1.8 & 87.47 $\pm$ 0.3\\
        &MNIST-Bin(N)  &  95.69$\pm$ 0.9&92.39 $\pm$ 0.5 &  87.48 $\pm$ 0.3 & 72.35 $\pm$ 2.2\\  \midrule    
    \end{tabular} 
    \vspace*{-0.3cm}
\end{table}\\
$\noindent$ \textit{\textbf{4.2.6 QFL performance under noisy channel:}}
Finally, we evaluated our proposed QFL model under noisy communication channels. We consider four noise communication channels as bit flip, phase flip, depolarizing and amplitude damping (Eq.~\ref{eq:bitflip} - Eq.~\ref{eq:ampdamp}) with varying noise intensities as 0.01, 0.05, 0.1 and 0.25. The results presented in Table~\ref{tab:3} illustrate the impact of noise intensity on MNIST-Bin and MNIST-Bin(N) for classification tasks. With bitflip noise, as the intensity increases, accuracy decreases from 97\% to 80\% due to information loss. Similar trends are observed with the other three noise channels. In case of noisy datasets, a rapid decrease in accuracy is observed from 94\% to 68\% in the case of a bitflip noisy channel. Similar results are observed across other noisy channels also. The boxplot shown in Figure~\ref{fig:fig7} illustrates the variation of QFL accuracy under various noise conditions as high fluctuations arise due to the randomness of noise.\\
$\noindent \blacksquare$ \textbf{Observations:} The primary observation is that quantum communication noise adversely affects the performance of NAC-QFL, with performance deteriorating as noise intensity increases. Additionally, in case of noisy datasets, the effect of noise multiplies and performance degrades rapidly with increase in noise intensity. 
\vspace*{-0.2cm}
\section{Conclusion}\label{sec:5}
The paper introduced a clustered quantum federated learning system for noisy quantum devices. It addresses the challenges of quantum noise, limited device capacity, and high communication costs. The system improved convergence through distributed task execution via circuit cutting and noise-aware device selection, and effectively mitigating noise's adverse impact on model accuracy. Experiment results demonstrated faster convergence, comparable performance, and reduced communication costs. We conclude that noise-aware quantum federated learning optimizes near-term quantum hardware utilization despite capacity, noise, and qubit decoherence limitations. Future research avenues include exploring methods to enhance the robustness of quantum federated learning systems against adversarial attacks and privacy breaches.
\bibliographystyle{ACM-Reference-Format}
\bibliography{paper}
\end{document}